\documentclass[prl,twocolumn,showpacs,preprintnumbers,superscriptaddress,amsmath,floatfix,amssymb,prl]{revtex4-1}
\usepackage[colorlinks=true]{hyperref}
\usepackage{graphicx}
\usepackage{epstopdf}
\newcommand{\comment}[1]{}

\newcommand{\lr}[1]{ \left( #1 \right) }
\newcommand{\lrs}[1]{ \left[ #1 \right] }

\newcommand{\vev}[1]{ \langle \, #1 \, \rangle }

\newcommand{\tr}{ {\rm Tr} \, }

\begin{document}
\sloppy
\preprint{ITEP-LAT/2010-01}

\title{Magnetic-Field-Induced insulator-conductor transition \\ in $SU\lr{2}$ quenched lattice gauge theory}

\author{P. V. Buividovich}
\affiliation{ITEP, B.~Cheremushkinskaya 25, Moscow, 117218, Russia}
\affiliation{JINR, Joliot-Curie 6, 141980 Dubna, Moscow region, Russia}

\author{M. N. Chernodub}\thanks{On leave from ITEP, Moscow, Russia.}
\affiliation{LMPT, CNRS UMR 6083, Universit\'e de Tours,
Parc de Grandmont, F37200, Tours, France}
\affiliation{DMPA, University of Gent, Krijgslaan 281, S9, B-9000 Gent, Belgium}

\author{D. E. Kharzeev}
\affiliation{Physics Department, Brookhaven National Laboratory, Upton, New York 11973-5000, USA}
\affiliation{Department of Physics, Yale University, New Haven, Connecticut 06520-8120, USA}

\author{T.Kalaydzhyan}
\affiliation{DESY Hamburg, Theory Group, Notkestrasse 85, D22607 Hamburg, Germany}
\affiliation{ITEP, B.~Cheremushkinskaya 25, Moscow, 117218, Russia}

\author{E. V. Luschevskaya}
\affiliation{ITEP, B.~Cheremushkinskaya 25, Moscow, 117218, Russia}
\affiliation{JINR, Joliot-Curie 6, 141980 Dubna, Moscow region, Russia}

\author{M. I. Polikarpov}
\affiliation{ITEP, B.~Cheremushkinskaya 25, Moscow, 117218, Russia}

\date{\today}
\begin{abstract}
We study the correlator of two vector currents in quenched $SU\lr{2}$ lattice gauge theory with a chirally invariant lattice Dirac operator with a constant external magnetic field. It is found that in the confinement phase the correlator of the components of the current parallel to the magnetic field decays much slower than in the absence of a magnetic field, while for other components the correlation length slightly decreases. We apply the maximal entropy method to extract the corresponding spectral function. In the limit of zero frequency this spectral function yields the electric conductivity of the quenched theory. We find that in the confinement phase the external magnetic field induces nonzero electric conductivity along the direction of the field, transforming the system from an insulator into an anisotropic conductor. In the deconfinement phase the conductivity does not exhibit any sizable dependence on the magnetic field.

\end{abstract}
\pacs{11.30.Rd; 12.38.Gc; 13.40.-f}
\maketitle

Recently, heavy ion experiments at  the BNL Relativistic Heavy Ion Collider (RHIC) have found an evidence \cite{ref:STAR} for the so-called chiral magnetic effect \cite{Kharzeev:2004ey} in quark-gluon plasma. The essence of the effect is the generation of electric current  along the direction of the external magnetic field in the background of topologically nontrivial gauge field configurations. Experimentally, the effect manifests itself as the dynamical enhancement of fluctuations in the numbers of charged hadrons emitted above and below the reaction plane in off-central heavy-ion collisions. Recently this effect has been studied also in lattice gauge theory, and the evidence for charge separation in magnetic field has been found \cite{Buividovich:09:7, Blum:09:1}. In \cite{Buividovich:09:7} it has been found that the fluctuations of the electric current along the magnetic field are strongly enhanced as compared to the fluctuations of current in the perpendicular directions. This conclusion was also confirmed by an analytical calculation in the instanton gas model \cite{Nam:09:1}. The result of \cite{Buividovich:09:7} on the difference of longitudinal and transverse electric current susceptibilities has been  reproduced later by an analytical calculation  \cite{Fukushima:2009ft}; the frequency dependence of the conductivity has also been evaluated -- for the weak coupling result, see \cite{Kharzeev:2009pj}.

A natural question to ask is whether this enhancement of current fluctuations
corresponds to a real flow of charge, or is just caused by short-lived quantum fluctuations. This question can be answered by studying the current-current correlation functions. The currents which correspond to a real transport of charged particles should have long-range correlations in time, while quantum fluctuations are typically characterized by a finite correlation time \cite{Kadanoff:63:1}. Recalling Green-Kubo relations, one can see that this property is intimately related to the electric conductivity - namely, the real transport of charged particles can occur only in conducting media. In this Letter we study the tensor of electric conductivity of the vacuum of quenched $SU\lr{2}$ lattice gauge theory in external magnetic field. We find that the magnetic field induces nonzero electric conductivity along its direction, transforming the confining vacuum from an insulator into an anisotropic conductor.

Electric conductivity can be extracted from the correlator of two vector currents $j_{i}\lr{x} = \bar{q}\lr{x} \gamma_{i} q\lr{x}$:
\begin{eqnarray}
G_{ij}\lr{\tau} = \int d^3 \vec{x} \vev{ j_{i}(\vec{0}, 0) j_{j}\lr{\vec{x}, \tau}  }
\label{corr_def}
\end{eqnarray}

 Following \cite{Aarts:07:1}, let us define the spectral function $\rho\lr{w}$ which corresponds to the correlator (\ref{corr_def})
\begin{eqnarray}
\label{spectral_def}
 G_{ij}\lr{\tau} & = & \int\nolimits_{0}^{+\infty} \frac{d w}{2 \pi}\, K\lr{w, \tau} \rho_{ij}\lr{w}, \\
\label{kernel_def}
 K\lr{w, \tau} & = & \frac{w}{2 T}  \, \frac{\cosh{\lr{w \lr{\tau - \frac{1}{2 T}}}}}
{\sinh{\lr{\frac{w}{2 T}}}},
\end{eqnarray}
where $T$ is the temperature. The Kubo formula for the electric conductivity then reads \cite{Kadanoff:63:1, Aarts:07:1}:
\begin{eqnarray}
\label{Kubo_formula}
\sigma_{ij} = \lim_{\omega \to 0} \frac{\rho_{ij}\lr{\omega}}{4 T}\,.
\end{eqnarray}
In the limit of the weak time-independent \emph{electric} field $E_{k}$, one has $\vev{j_{i}} = \sigma_{ik} E_{k}$. Thus electric conductivity is related to the behavior of the spectral function at small frequencies. If there is a gap in the spectrum so that $\rho_{ij}\lr{w} = 0$ for $w < w_{c}$, electric conductivity is zero and $G_{ij}\lr{\tau} \sim \cosh\lr{w_{c} \lr{\tau - \frac{1}{2 T}}}$. On the other hand, if $\rho\lr{w}$ is not zero near $w = 0$, one can expect slow nonexponential decay of $G_{ij}\lr{\tau}$.

To measure the correlator (\ref{corr_def}), we perform lattice Monte-Carlo simulations of quenched $SU\lr{2}$ lattice gauge theory. Since quark chirality is very important for magnetic effects in non-Abelian gauge theories \cite{Kharzeev:2004ey}, we use the overlap lattice Dirac operator $\mathcal{D}$ with exact chiral symmetry \cite{Neuberger:98:1} to measure the vector currents. We consider the two-current correlator in the meson channel, which is represented in terms of Dirac propagators in fixed Abelian and non-Abelian gauge fields and is then averaged over an equilibrium ensemble of non-Abelian gauge fields $A_\mu$:
\begin{eqnarray}
\label{four_fermion_vev}
& & \vev{\bar{q}\lr{x} \gamma_{i} q\lr{x} \, \bar{q}\lr{y} \gamma_{j} q\lr{y}}
\nonumber \\ & & =
\int \mathcal{D}A_{\mu}\,e^{-S_{YM}\lrs{A_{\mu}}}\, \tr\lr{\frac{1}{\mathcal{D} + m} \, \gamma_{i} \, \frac{1}{\mathcal{D} + m} \, \gamma_{j}},
\qquad
\end{eqnarray}
where $S_{YM}\lrs{A_{\mu}}$ is the lattice action for gluons $A_{\mu}$.  A uniform magnetic field is added to the Dirac operator by substituting $su\lr{2}$-valued vector potential $A_{\mu}$ with $u\lr{2}$-valued one $A_{\mu \, ij} \rightarrow A_{\mu \, ij} + 1/2\: F_{\mu\nu}\: x_{\nu} \delta_{ij}$. In order to account for periodic boundary conditions we introduce an additional twist for fermions \cite{Wiese:08:1, Buividovich:09:7}. The quark mass is fixed in lattice units at a small value $a m = 0.01$. Previous studies of mesonic correlation functions with an overlap Dirac operator indicate that the vector current correlator depends very weakly on quark mass~\cite{Babich:05:1}.

Strictly speaking, the correlator (\ref{four_fermion_vev}) corresponds to the correlator of
charged currents, for example $\bar{u} \gamma_{\mu} d$. The correlator of neutral
currents considered in our Letter, $j_{\mu} = \bar{d} \gamma_{\mu} d$, should also contain the disconnected part.
This part is quite intricate for an accurate numerical
treatment. We have roughly estimated its contribution by inverting the
Dirac operator on a subspace spanned on some small number $M \sim 30$ of
the lowest Dirac eigenmodes, as in \cite{Hasenfratz:02:2}. It turned out
that this part of the full neutral current correlator behaves
similarly to the connected one (\ref{four_fermion_vev}). We do not reproduce these
estimates here due to uncontrollable systematic errors~\cite{Hasenfratz:02:2}.

We use the tadpole-improved Wilson-Symanzik action (see, e.g., Eq.~(1) in \cite{Luschevskaya:08:1}). 
For inversion, we use a Gaussian source with radius $r = 1.0$ in lattice units in both spatial and time directions and a point sink (that is, quark position is smeared over a Gaussian profile). We have found that such smearing significantly improves the convergence of the maximal entropy method \cite{Aarts:07:1, Gupta:04:1, Asakawa:01:1} at small lattice sizes, while the value of the conductivity is practically unaffected.
Our lattice parameters are summarized in Table \ref{tab:params}. A uniform magnetic field is introduced into the Dirac operator as described in \cite{Buividovich:09:7}. In order to obtain the Dirac propagator, we implement the shifted unitary minimal residue method of Ref.~\cite{Borici:06}.

\begin{table}[ht]
  \centering
  \begin{tabular}{|c|c|c|c|c|}
  \hline
  $\beta$  & a, \mbox{fm} & $N_{s}^{3} \times N_{t}$ & $T/T_c$ & \#conf \\
  \hline
  3.2810   & 0.102 & $14^3 \times 14$         & 0.43    & 30 \\
  3.2810   & 0.102 & $16^3 \times 16$         & 0.38    & 30 \\
  3.3555   & 0.089 & $16^3 \times 16$         & 0.43    & 30 \\
  3.3250   & 0.095 & $16^3 \times 6$          & 1.12    & 30 \\
  \hline
  \end{tabular}
  \caption{Lattice parameters used in our simulations. The critical temperature of the deconfinement phase transition in quenched $SU\lr{2}$ gauge theory is $T_{c} = 313.(3) \, \mbox{MeV}$~\cite{Bornyakov:07:1}.}
  \label{tab:params}
\end{table}

It is clear that since the magnetic field is parallel to the $z$ axis, the principal axes of the tensor $\sigma_{ij}\lr{\tau}$ will be the $x$, $y$ and $z$ axes and it is sufficient to consider only the diagonal components $\sigma_{ii}$ (no summation over $i$=$x$, $y$,~$z$).

\begin{figure*}[ht]
  \includegraphics[width=5.8cm, angle=-90]{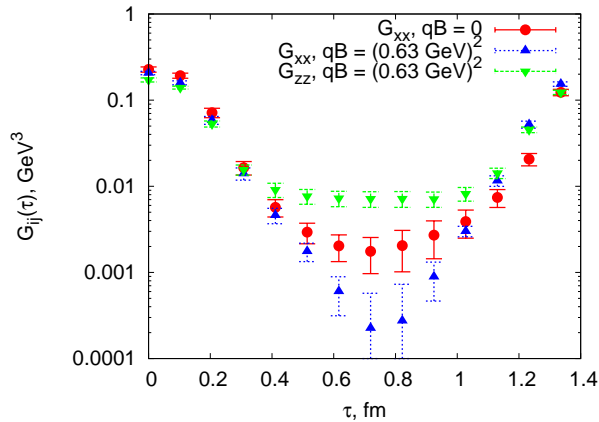} \hskip 5mm
  \includegraphics[width=5.8cm, angle=-90]{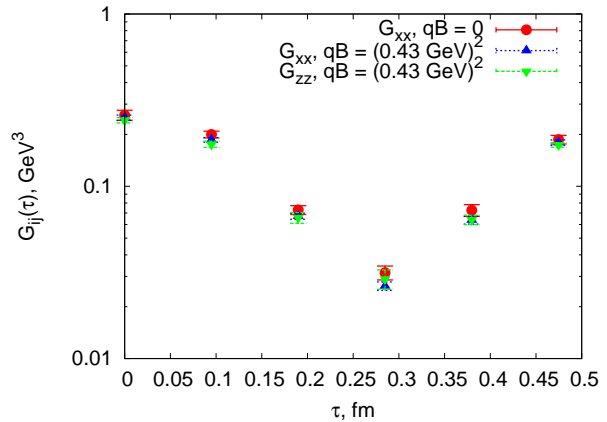}
\vskip -1mm
  \caption{The correlator (\ref{corr_def}) in the confinement (left) and in the deconfinement phases (right) at $T = 350\ {\rm MeV}$.}
  \label{fig:corrs}
\end{figure*}

We plot some correlators at different temperatures and magnetic fields on Fig. \ref{fig:corrs}. The data are for the $14^4$ lattice with spacing $a = 0.102 \, \mbox{fm}$ (left) and for the $16^3 \times 6$ lattice with spacing $a = 0.095 \, \mbox{fm}$ (right). For the latter lattice the temperature is $T = 350\, \mbox{MeV} = 1.12\,T_c$ and the theory is in the deconfinement phase.
In the quenched theory the critical temperature of the deconfinement transition is not affected by the magnetic field.
The temperature $T=1.12$ corresponds to the chirally restored phase.

One can see that without the magnetic field the correlators decay quickly in the confinement phase. In the deconfinement phase the decay is significantly slower for all $G_{ii}\lr{\tau}$. When we switch on a magnetic field with the strength $q B = \lr{0.63\ {\rm GeV}}^2$, in the confinement phase the correlator $G_{zz}\lr{\tau}$ decays much slower and is significantly larger than zero for all $\tau$, much like in the deconfinement phase. In contrast, the correlators for the perpendicular components of the current $G_{xx}\lr{\tau}$ and $G_{yy}\lr{\tau}$ decay somewhat quicker than in the zero field case. In the deconfinement phase all the correlators are practically unaffected by the magnetic field.

\begin{figure}[ht]
\vskip -5mm
  \includegraphics[width=5.8cm, angle=-90]{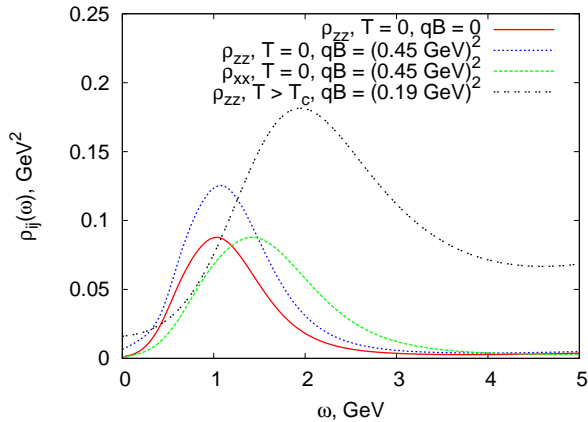}
\vskip -1mm
  \caption{Spectral functions $\rho_{ij}\lr{w}$ in the confinement and deconfinement phases.}
  \label{fig:spectral}
\end{figure}

We now apply the Maximal Entropy Method \cite{Aarts:07:1, Gupta:04:1, Asakawa:01:1} to extract the spectral functions (\ref{spectral_def}) from the correlators (\ref{corr_def}). Our analysis is similar to that of Refs.~\cite{Aarts:07:1, Asakawa:01:1}. We used the model with the default guess $\bar{m}\lr{w} = \bar{m}_{0}\lr{b + a w}$ \cite{Aarts:07:1}. Some spectral functions at different temperatures and magnetic fields are plotted on Fig. \ref{fig:spectral}.

In the confinement phase and in the absence of magnetic field, the spectral function has a distinct peak near $w \approx 1 \, \mbox{GeV}$, which corresponds to the mass of the $\rho$ meson in quenched $SU\lr{2}$ lattice gauge theory \cite{Asakawa:01:1, Hasenfratz:02:2}. The width of this peak in quenched approximation is a lattice artifact \cite{Asakawa:01:1}, and should decrease for finer and larger lattices. The spectral function in the limit of zero frequency, $\rho_{ij}\lr{0}$, is equal to zero within error range. This indicates that in the absence of an external magnetic field the vacuum of quenched QCD is an insulator, in agreement with the results of \cite{Asakawa:01:1, Gupta:04:1}. When the external magnetic field is applied, the peak grows and the spectral function becomes nonzero in the limit of zero frequency. For other components of $\rho_{ij}\lr{w}$ nothing changes qualitatively, but the peak which corresponds to the $\rho$-meson becomes somewhat smaller and shifts slightly to larger $w$. The conductivity stays equal to zero within the error range. Thus when the external magnetic field is applied to the quenched vacuum of the $SU\lr{2}$ lattice gauge theory, the vacuum acquires nonzero conductivity, but only in the direction of the magnetic field.

In the deconfinement phase at zero magnetic field, the spectral function is nonzero at $w = 0$ and has a smooth peak near $w \approx 2 \, {\rm GeV}$. Thus quenched $SU\lr{2}$ lattice gauge theory is a conductor above the deconfinement phase transition \cite{Aarts:07:1, Gupta:04:1}. Since the shape of the correlator $G_{ij}\lr{\tau}$ is practically unaffected by the magnetic field, the spectral function $\rho_{ij}\lr{w}$ and the conductivity $\sigma_{ij}$ do not depend on the magnetic field.

\begin{figure}[ht]
\vskip -5mm
  \includegraphics[width=5.8cm, angle=-90]{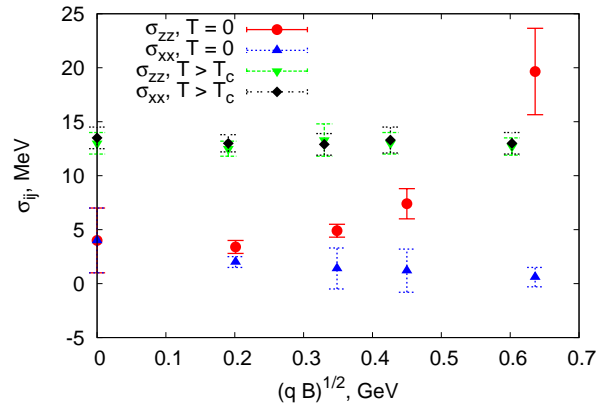}
\vskip -1mm
  \caption{Electric conductivity of quenched QCD as a function of an external magnetic field at different temperatures. The points for $\sigma_{zz}$ and $\sigma_{xx}$ at $T{>} T_c$ coincide within the errors.}
  \label{fig:conductivity}
\end{figure}

The electric conductivity $\sigma_{ij}$ as a function of external magnetic field is plotted on Fig. \ref{fig:conductivity} for the confinement and deconfinement phases. In the deconfinement phase the temperature is $T = 350 \, {\rm MeV}$. The value of the conductivity was extracted from the value of the spectral function at $w = 0$ using (\ref{Kubo_formula}). In the confinement phase and at zero magnetic field the conductivity is zero within the error range. As the magnetic field is turned on, the conductivity $\sigma_{zz}$ in the direction of the magnetic field grows, while all other components of $\sigma_{ij}$ remain equal to zero within error range. In the deconfinement phase the conductivity is isotropic and is practically independent of the magnetic field. One can not exclude, of course, that there is a weak anisotropy, which cannot be seen at the small number of configurations that we have. It should be also noted that in our simulations the value of conductivity $\sigma = 15 \pm 2 \, {\rm MeV}$ at $T = 350 \, {\rm MeV} > T_c$ is still much smaller than the results obtained in \cite{Gupta:04:1, Aarts:07:1} in quenched $SU\lr{3}$ lattice gauge theory with light staggered fermions. This difference is likely to be an artifact of a quenched theory, since in this case different probes of the confinement-deconfinement phase transition might give different transition temperatures. In particular, while in quenched $SU\lr{2}$ lattice gauge theory the Polyakov loop goes to zero at $T_{c} = 313.(3) \, {\rm MeV}$  \cite{Bornyakov:07:1}, the chiral condensate is not zero above this temperature \cite{Buividovich:08:5}. The situation might be similar for the insulator-conductor transition, which in the quenched case might be replaced by a soft crossover with much smaller conductivity at $T > T_c$.

\begin{figure}[!ht]
  \includegraphics[width=5.5cm, angle=-90]{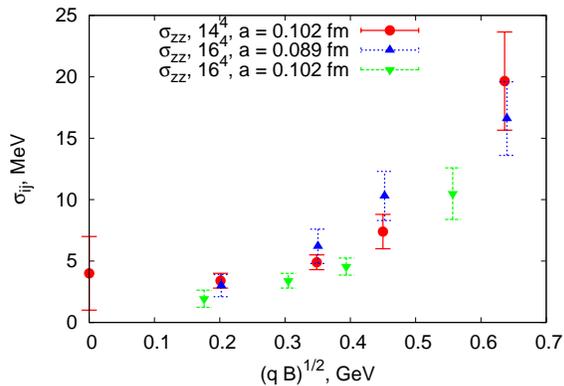}
  \caption{Electric conductivity in the direction of external magnetic field $\sigma_{zz}$ for different lattice parameters.}
\vskip -1mm
  \label{fig:conductivity_latt}
\end{figure}

The transport coefficients typically have rather strong dependence on lattice parameters. To ensure that the nonzero conductivity is not a finite-volume artifact, we have also performed the simulations at different lattice volumes and lattice spacings (see Table \ref{tab:params}). The values of conductivity $\sigma_{zz}$ for different lattice parameters are plotted on Fig. \ref{fig:conductivity_latt}. One can see that as we go to finer and larger lattices, the conductivity does not change within statistical errors.

We conclude that a strong magnetic field can induce nonzero electric conductivity of the vacuum of quenched non-Abelian lattice gauge theory along the direction of the field, turning it into an anisotropic conductor. This effect may be called ``electric rupture facilitated by  magnetic field''; it may originate from the interplay of gluon field topology and an increase in the quark zero mode density due to the presence of magnetic field.  It can be interesting to investigate whether there is some critical value of the magnetic field at which the conductivity becomes nonzero. Transitions of this type are known in condensed-matter physics \cite{Jonker}. In contrast, in the deconfinement phase the vacuum is an isotropic conductor, and the value of the conductivity is practically independent of the magnetic field.
Thus, if a strong magnetic field generates an electric current via the chiral magnetic effect in a CP-odd background, then the sufficiently strong field would guarantee that the charge will propagate through the media due to finite electric conductivity in the both phases.

Finally, let us comment on possible experimental consequences of the phenomenon described above. The expectation value $\vev{ j_{k}\lr{x} j_{l}\lr{y}}$ is related to the polarization of soft photons and to the angular distribution of soft photons and dilepton pairs emitted in the collision process \cite{Gupta:04:1,McLerran:85:1}. One can therefore expect an enhancement of  production rate in the direction perpendicular to the reaction plane, possibly resulting in a negative elliptic flow for soft photons and dilepton pairs.

\begin{acknowledgments}
The authors are grateful to A. S. Gorsky, A. Krikun, V. I. Shevchenko, O. V. Teryaev and V. I. Zakharov for interesting and useful discussions. This work was partly supported by Grants RFBR No. 08-02-00661-a, 09-02-00338-a and DFG-RFBR 436 RUS, a grant for scientific schools No. NSh-679.2008.2, and by the Russian Federal Agency for Nuclear Power.  The work of D.K. was supported by Contract No.DE-AC02-98CH10886 with the U.S. Department of Energy. P. B. was partially supported by personal grants from the Dynasty foundation and from the FAIR-Russia Research Center (FRRC). The work of M.N.C. was partially supported by the French Agence Nationale de la Recherche project ANR-09-JCJC ``HYPERMAG''. The calculations were partially done on the MVS 50K at Moscow Joint Supercomputer Center.
\end{acknowledgments}

\end{document}